# DETERMINING OPTIMAL TRADING RULES WITHOUT BACKTESTING

Peter P. Carr
*Global Head of Market Modeling – Morgan Stanley*
*Mathematics Department – Courant Institute of Mathematical Sciences*
pcarr@nyc.rr.com
www.math.nyu.edu/research/carrp

Marcos López de Prado
*Senior Managing Director – Guggenheim Partners*
*Research Affiliate - Lawrence Berkeley National Laboratory*
lopezdeprado@lbl.gov
www.QuantResearch.info

# DETERMINING OPTIMAL TRADING RULES WITHOUT BACKTESTING

### ABSTRACT

Calibrating a trading rule using a historical simulation (also called backtest) contributes to backtest overfitting, which in turn leads to underperformance. In this paper we propose a procedure for determining the *optimal trading rule* (OTR) without running alternative model configurations through a backtest engine. We present empirical evidence of the existence of such optimal solutions for the case of prices following a discrete Ornstein-Uhlenbeck process, and show how they can be computed numerically. Although we do not derive a closed-form solution for the calculation of OTRs, we conjecture its existence on the basis of the empirical evidence presented.

# 1.- INTRODUCTION

Investment strategies can be defined as logical arguments that postulate the existence of a market inefficiency. Some strategies use econometric arguments to forecast financial variables such as GDP or inflation; other strategies use fundamental and accounting information to price securities; or search for arbitrage-like opportunities in the pricing of derivatives products, etc. For instance, suppose that banking corporations tend to sell off-the-run bonds two days before U.S. Treasury auctions, in order to reserve balance sheet for the new "paper". One could monetize on that knowledge by selling off-the-run bonds three days before auctions. But how? Each investment strategy requires an implementation tactic, often referred to as *trading rules*.

There are dozens of hedge fund styles, each running dozens of unique investment strategies. While strategies can be very heterogeneous in nature, tactics are relatively homogeneous. Trading rules provide the algorithm that must be followed to enter and exit a position. For example, a position will be entered when the strategy's signal reaches a certain value. Conditions for exiting a position are often defined through thresholds for profit-taking and stop-losses. These entry and exit rules rely on parameters that are usually calibrated via historical simulations. This practice leads to the problem of *backtest overfitting*, because these parameters target specific observations in-sample, to the point that the investment strategy is so attached to the past that becomes unfit for the future.

An important clarification is that we are interested in the exit corridor conditions that maximize performance. In other words, the position already exists and the question is how to exit it optimally. This is the dilemma often faced by execution traders, and it should not be mistaken with the determination of entry and exit thresholds for some underlying instrument. For a study of that alternative question, see Bertram [2009].

Bailey et al. [2013, 2014] discuss the problem of backtest overfitting, and provide methods to determine to what extent a simulated performance may be inflated due the overfitting. While assessing the probability of backtest overfitting is a useful tool to discard superfluous investment strategies, it would be better to avoid the risk of overfitting, at least in the context of calibrating a trading rule. In theory this could be accomplished by estimating the optimal parameters for the trading rule directly from the data, rather than engaging in historical simulations. This is the approach we take in this paper. Using the entire historical sample, we will characterize the stochastic process that generates the observed stream of returns, and derive the optimal values for the trading rule's parameters without requiring a historical simulation.

The rest of the study is organized as follows: Section 2 defines a trading rule, sets its characterization and introduces the problem of overfitting in the context of a trading rule's calibration. Section 3 describes our framework for calibrating a trading rule. Section 4 illustrates how to determine optimal trading rules (OTRs) numerically. Section 5 summarizes our conclusions. The appendices present an implementation in Python of our experiments.

## 2.- THE PROBLEM

Suppose an investment strategy $S$ that invests in $i = 1, \ldots I$ opportunities or bets. At each opportunity $i$, $S$ takes a position of $m_i$ units of security $X$, where $m_i \in (-\infty, \infty)$. The transaction that entered such opportunity was priced at a value $m_i P_{i,0}$, where $P_{i,0}$ is the average price per unit at which the $m_i$ securities were transacted. As other market participants transact security $X$, we can mark-to-market (MtM) the value of that opportunity $i$ after $t$ observed transactions as $m_i P_{i,t}$. This represents the value of opportunity $i$ if it were liquidated at the price observed in the market after $t$ transactions. Accordingly, we can compute the MtM profit/loss of opportunity $i$ after $t$ transactions as $\pi_{i,t} = m_i(P_{i,t} - P_{i,0})$.

A standard trading rule provides the logic for exiting opportunity $i$ at $t = T_i$. This occurs when one of two conditions is verified:
- $\pi_{i,T_i} \geq \overline{\pi_i}$, where $\overline{\pi_i} > 0$ is the profit-taking threshold for opportunity $i$.
- $\pi_{i,T_i} \leq \underline{\pi_i}$, where $\underline{\pi_i} < 0$ is the stop-loss threshold for opportunity $i$.

Because $\underline{\pi_i} < \overline{\pi_i}$, only one of the two exit conditions can trigger the exit from opportunity $i$. Assuming that opportunity $i$ can be exited at $T_i$, its final profit/loss is $\pi_{i,T_i}$. At the onset of each opportunity, the goal is to realize an expected profit $E_0[\pi_{i,T_i}] = m_i(E_0[P_{i,T_i}] - P_{i,0})$, where $E_0[P_{i,T_i}]$ is the forecasted price and $P_{i,0}$ is the entry level of opportunity $i$.

*<u>DEFINITION 1 (Trading Rule)</u>: A trading rule for strategy $S$ is defined by the set of parameters $R := \{(\underline{\pi_i}, \overline{\pi_i})\}, i = 1, \ldots I$.*

One way to calibrate the trading rule is to:
1. Define a set of alternative values of $R$, $\Omega := \{R\}$.
2. Simulate historically (also called backtest) the performance of $S$ under alternative values of $R \in \Omega$.
3. Select the optimal $R^*$.

More formally:

$$R^* = \arg\max_{R \in \Omega} SR_R$$
$$SR_R = \frac{E[\pi_{i,T_i}|R]}{\sigma[\pi_{i,T_i}|R]} \quad (1)$$

where $E[.]$ and $\sigma[.]$ are respectively the expected value and standard deviation of $\pi_{i,T_i}$, conditional on trading rule $R$, over $i = 1, \ldots I$. In other words, Eq. (1) is maximizing the Sharpe ratio of $S$ on $I$ opportunities over the space of alternative trading rules $R$ (see Bailey and López de Prado [2012] for a definition and analysis of the Sharpe ratio). Because we count with $2I$ variables to maximize $SR_R$ over a sample of size $I$, it is easy to overfit $R$. A trivial overfit occurs when each pair $\left(\underline{\pi_i}, \overline{\pi_i}\right)$ targets the specific opportunity $i$. Bailey et al. [2013] provide a rigorous definition of backtest overfitting, which can be applied to our study of trading rules as follows.

*DEFINITION 2 (Overfit Trading Rule):* $R^*$ is overfit if
$$E\left[\frac{E[\pi_{j,T_j}|R^*]}{\sigma[\pi_{j,T_j}|R^*]}\right] < Me_\Omega\left[E\left[\frac{E[\pi_{j,T_j}|R]}{\sigma[\pi_{j,T_j}|R]}\right]\right], \text{ where } j = I+1, ... J.$$

Intuitively, an optimal in-sample (IS) trading rule $R^*$ is overfit when it is expected to underperform the median of alternative trading rules $R \in \Omega$ out-of-sample (OOS). Bailey et al. [2014] argue that it is hard not to overfit a backtest, particularly when there are free variables able to target specific observations IS, or the number of elements in $\Omega$ is large. A trading rule introduces such free variables, because $R^*$ can be determined independently from $S$. The outcome is that the backtest profits from random noise IS, making $R^*$ unfit for OOS opportunities. Those same authors show that overfitting leads to negative performance OOS when $\Delta\pi_{i,t}$ exhibits serial dependence. While those authors provide a useful method to evaluate to what extent a backtest has been overfit, it would be convenient to avoid this problem in the first place.[1] To that aim we dedicate the following section.

**3.- OUR FRAMEWORK**
Until now we have not characterized the stochastic process from which observations $\pi_{i,t}$ are drawn. We are interested in providing an OTR for those scenarios where overfitting would be most damaging, such as when $\pi_{i,t}$ exhibits serial correlation. In particular, suppose a discrete Ornstein-Uhlenbeck (O-U) process on prices

$$P_{i,t} = (1-\varphi)E_0[P_{i,T_i}] + \varphi P_{i,t-1} + \sigma\varepsilon_{i,t} \qquad (2)$$

such that the random shocks are IID distributed $\varepsilon_{i,t} \sim N(0,1)$. The seed value for this process is $P_{i,0}$, the level targeted by opportunity $i$ is $E_0[P_{i,T_i}]$, and $\varphi$ determines the speed at which $P_{i,0}$ converges towards $E_0[P_{i,T_i}]$. Because $\pi_{i,t} = m_i(P_{i,t} - P_{i,0})$, Eq. (2) implies that the performance of opportunity $i$ is characterized by the process

$$\frac{1}{m_i}\pi_{i,t} = (1-\varphi)E_0[P_{i,T_i}] - P_{i,0} + \varphi P_{i,t-1} + \sigma\varepsilon_{i,t} \qquad (3)$$

From the proof to Proposition 4 in Bailey and López de Prado [2013], it can be shown that the distribution of the process specified in Eq. (2) has a closed-form in

$$\pi_{i,t} \sim N\left( m_i \left( (1-\varphi)E_0[P_{i,T_i}]\sum_{j=0}^{t-1}\varphi^j - P_{i,0} \right), m_i^2\sigma^2 \sum_{j=0}^{t-1}\varphi^{2j} \right) \quad (4)$$

and a necessary and sufficient condition for its stationarity is that $\varphi \in (-1,1)$. Given a set of input parameters $\{\sigma, \varphi\}$ and initial conditions $\{P_{i,0}, E_0[P_{i,T_i}]\}$ associated with opportunity $i$, is there an OTR $R_i^* := (\underline{\pi_i}, \overline{\pi_i})$? Similarly, should strategy $S$ predict a profit target $\overline{\pi_i}$, can we compute the optimal stop-loss $\underline{\pi_i}$ given the input values $\{\sigma, \varphi\}$? If the answer to these questions is affirmative, no backtest would be needed in order to determine $R_i^*$, thus avoiding the problem of overfitting the trading rule. In the next section we will show how to answer these questions experimentally.

---
[1] The strategy may still be the result of backtest overfitting, but at least the trading rule would not have contributed to that problem.

## 4.- NUMERICAL DETERMINATION OF OTRs

In the previous section we used an O-U specification to characterize the stochastic process generating the returns of strategy $S$. In this section we will present a procedure to derive numerically the OTR for any specification in general, and the O-U specification in particular.

<u>STEP 1</u>: We estimate the input parameters $\{\sigma, \varphi\}$, by linearizing Eq. (2) as:

$$P_{i,t} = E_0[P_{i,T_i}] + \varphi(P_{i,t-1} - E_0[P_{i,T_i}]) + \xi_t \quad (5)$$

We can then form vectors X and Y by sequencing opportunities:

$$X = \begin{bmatrix} P_{0,0} - E_0[P_{0,T_0}] \\ P_{0,1} - E_0[P_{0,T_0}] \\ \dots \\ P_{0,T-1} - E_0[P_{0,T_0}] \\ \dots \\ P_{I,0} - E_0[P_{I,T_I}] \\ \dots \\ P_{I,T-1} - E_0[P_{I,T_I}] \end{bmatrix}; Y = \begin{bmatrix} P_{0,1} \\ P_{0,2} \\ \dots \\ P_{0,T} \\ \dots \\ P_{I,1} \\ \dots \\ P_{I,T} \end{bmatrix}; Z = \begin{bmatrix} E_0[P_{0,T_0}] \\ E_0[P_{0,T_0}] \\ \dots \\ E_0[P_{0,T_0}] \\ \dots \\ E_0[P_{I,T_I}] \\ \dots \\ E_0[P_{I,T_I}] \end{bmatrix} \quad (6)$$

Applying OLS on Eq. (5), we can estimate the original O-U parameters as,

$$\begin{aligned} \hat{\varphi} &= \frac{cov[Y,X]}{cov[X,X]} \\ \hat{\xi}_t &= Y - Z - \hat{\varphi}X \\ \hat{\sigma} &= \sqrt{cov(\hat{\xi}_t, \hat{\xi}_t)} \end{aligned} \quad (7)$$

STEP 2: We construct a mesh of stop-loss and profit-taking pairs, $\left(\underline{\pi_i}, \overline{\pi_i}\right)$. For example, a Cartesian product of $\underline{\pi_i} = \left\{0, -\frac{1}{2}\sigma, -\sigma, \ldots, -10\sigma\right\}$ and $\overline{\pi_i} = \left\{0, \frac{1}{2}\sigma, \sigma, \ldots, 10\sigma\right\}$ give us 21x21 nodes, each constituting an alternative trading rule $R_i$.

STEP 3: We generate a large number of paths (e.g., 100,000) for $\pi_{i,t}$ applying our estimates $\{\hat{\sigma}, \hat{\varphi}\}$. As seed values, we use the observed initial conditions $\{P_{i,0}, E_0[P_{i,T_i}]\}$ associated with an opportunity $i$. Because a position cannot be held for an unlimited period of time, we can impose a maximum holding period (e.g., 100 observations) at which point the position is exited even though $\underline{\pi_i} \leq \pi_{i,100} \leq \overline{\pi_i}$.

STEP 4: We apply the 100,000 paths generated in Step 3 on each node of the 21x21 mesh $\left(\underline{\pi_i}, \overline{\pi_i}\right)$ generated in Step 2. For each node, we apply the stop-loss and profit-taking logic, giving us 100,000 values of $\pi_{i,T_i}$. Likewise, for each node we compute the Sharpe ratio associated with that trading rule as described in Eq. (1) (see Bailey and López de Prado [2012] for a study of the confidence bands of the Sharpe ratio estimator). This result can be used in two different ways (Steps 5a, Step 5b and 5c):

STEP 5a: We determine the pair $\left(\underline{\pi_i}, \overline{\pi_i}\right)$ within the mesh of trading rules that is optimal, given the input parameters $\{\hat{\sigma}, \hat{\varphi}\}$ and the observed initial conditions $\{P_{i,0}, E_0[P_{i,T_i}]\}$.

STEP 5b: If strategy $S$ provides a profit target $\overline{\pi_i}$ for a particular opportunity $i$, we can use that information in conjunction with the results in Step 4 to determine the optimal stop-loss, $\underline{\pi_i}$.

STEP 5c: If the trader has a maximum stop-loss $\underline{\pi_i}$ imposed by the fund's management, we can use that information in conjunction with the results in Step 4 to determine the optimal profit taking $\overline{\pi_i}$ within the range of stop-losses $\left[0, \underline{\pi_i}\right]$.

Bailey et al. [2013a] proof that the half-life of the process in Eq. (2) is $\tau = -\frac{Ln[2]}{Ln[\varphi]}$, which implies the additional constraint $\varphi \in (0,1)$. From that result, we can determine the value of $\varphi$ associated with a certain half-life $\tau$ as $\varphi = 2^{-1/\tau}$.

Appendix 1 implements this procedure. Table 1 lists the combinations analyzed in this study. OTR $R_i^* := \left(\underline{\pi_i}, \overline{\pi_i}\right)$ is computed per unit held ($m_i = 1$), since other values of $m_i$ would simply re-scale performance. Although different values for these input parameters would render different numerical results, the combinations applied allow us to analyze the most general cases.

[TABLE 1 HERE]

In the following figures, we have plotted the non-annualized Sharpe ratios that result from various combinations of profit-taking and stop-loss. We have omitted the negative sign in the y-axis (stop-losses) for simplicity. Sharpe ratios are represented with different scales of colors (green indicating better performance and red worse performance), in a format known as a "heat-map".

**4.1.- CASES WITH ZERO LONG-RUN EQUILIBRIUM**
These cases are consistent with the business of market makers, who provide liquidity under the assumption that prices follow a martingale. The smaller $\tau$, the smaller is the autoregressive coefficient ($\varphi = 2^{-1/\tau}$). A small autoregressive coefficient in conjunction with a zero expected profit has the effect that most of the pairs $\left(\underline{\pi_i}, \overline{\pi_i}\right)$ deliver a zero performance.

Figure 1 shows the heat-map for the parameter combination $\{\mu, \tau, \sigma\} = \{0, 5, 1\}$. The half-life is so small that performance is maximized in a narrow range of combinations of small profit-taking with large stop-losses. In other words, the optimal trading rule is to hold an inventory long enough until a small profit arises, even at the expense of experiencing 5 or 7-fold losses. Sharpe ratios are high, reaching levels of around 3.2. This is in fact what many market-makers do in practice, and is consistent with the "asymmetric payoff dilemma" described in Easley et al. [2011]. The worst possible trading rule in this setting would be to combine a short stop-loss with large profit-taking threshold, a situation that market-makers avoid in practice. Performance is closest to neutral in the diagonal of the mesh, where profit-taking and stop-losses are symmetric.

[FIGURE 1 HERE]

Figure 2 shows that, if we increase $\tau$ from 5 to 10, the areas of highest and lowest performance spread over the mesh of pairs $\left(\underline{\pi_i}, \overline{\pi_i}\right)$, while the Sharpe ratios decrease. This is because, as the half-life increases, so that the magnitude of the autoregressive coefficient (recall that $\varphi = 2^{-1/\tau}$), thus approaching the process to a random walk.

[FIGURE 2 HERE]

In Figure 3, $\tau = 25$, which again spreads the areas of highest and lowest performance while reducing the Sharpe ratio. Figures 4 ($\tau = 50$) and 5 ($\tau = 100$) continue that progression. Eventually, as $\varphi \to 1$, there are no recognizable areas where performance can be maximized.

[FIGURE 3 HERE]

[FIGURE 4 HERE]

[FIGURE 5 HERE]

Calibrating a trading rule on a random walk through historical simulations would lead to backtest overfitting, because a random combination of profit-taking and stop-loss that happened to maximize Sharpe ratio would be selected. Our procedure prevents overfitting by recognizing that performance exhibits no consistent pattern, indicating that there is no optimal trading rule.

**4.2.- CASES WITH POSITIVE LONG-RUN EQUILIBRIUM**
These cases are consistent with the business of a position-taker, such as a hedge fund or asset manager. Figure 6 shows the results for the parameter combination $\{\mu, \tau, \sigma\} = \{5,5,1\}$. Because positions tend to make money, the optimal profit-taking is higher than in the previous cases, centered around 6, with stop-losses that range between 4 and 10. The region of the optimal trading rule takes a characteristic rectangular shape, as a result of combining a wide stop-loss range with a narrower profit-taking range. Performance is highest across all experiments, with Sharpe ratios of around 12.

[FIGURE 6 HERE]

In Figure 7, we have increased the half-life from $\tau = 5$ to $\tau = 10$. Now the optimal performance is achieved at a profit-taking centered around 5, with stop-losses that range between 7 and 10. The range of optimal profit-taking is wider, while the range of optimal stop-losses narrows, shaping the former rectangular area closer to a square. Again, a larger half-life brings the process closer to a random walk, and therefore performance is now relatively lower than before, with Sharpe ratios of around 9.

[FIGURE 7 HERE]

In Figure 8, we have made $\tau = 25$. The optimal profit-taking is now centered around 3, while the optimal stop-losses range between 9 and 10. The previous square area of optimal performance has given way to a semi-circle of small profit-taking with large stop-loss thresholds. Again we see a deterioration of performance, with Sharpe ratios of 2.7.

[FIGURE 8 HERE]

In Figure 9, half-life raises to $\tau = 50$. As a result, the region of optimal performance spreads, while Sharpe ratios continue to fall to 0.8. This is the same effect we observed in the case of zero long-run equilibrium (Section 4.1), with the difference that because now $\mu > 0$ there is no symmetric area of worst performance.

[FIGURE 9 HERE]

In Figure 10, we appreciate that $\tau = 100$ leads to the natural conclusion of the trend described above. The process is now so close to a random walk that the Maximum Sharpe ratio now is a mere 0.32.

[FIGURE 10 HERE]

We can observe a similar pattern in Figures 11 to 15, where $\mu = 10$ and $\tau$ is progressively increased from 5 to 10, 25, 50 and 100.

**4.3.- CASES WITH NEGATIVE LONG-RUN EQUILIBRIUM**
A rational market participant would not initiate a position under the assumption that a loss is the expected outcome. However, if a trader recognizes that losses are the expected outcome of a pre-existing position, she still needs a strategy to trade it away while minimizing such losses.

We have obtained Figure 16 as a result of applying parameters $\{\mu, \tau, \sigma\} = \{-5,5,1\}$. If we compare Figure 16 with Figure 6, it appears as if one is a rotated complementary of the other. Figure 6 resembles a rotated photographic negative of Figure 16. The reason is, that the profit in Figure 6 is translated into a loss in Figure 16, and the loss in Figure 6 is translated into a profit in Figure 16. One case is an image of the other, just as a gambler's loss is the house's gain.

As expected, Sharpe ratios are negative, with a worst performance region centered around the stop-loss of 6, and profit-taking threshold that range between 4 and 10. Now the rectangular shape does not correspond to region of best performance, but to a region of worst performance, with Sharpe ratios of around -12.

[FIGURE 16 HERE]

In Figure 17, $\tau = 10$, and now the proximity to a random walk plays in our favor. The region of worst performance spreads out, and the rectangular area becomes a square. Performance becomes less negative, with Sharpe ratios of about -9.

[FIGURE 17 HERE]

This familiar progression can be appreciated in Figures 18-20, as $\tau$ is raised to 25, 50 and 100. Again, as the process approaches a random walk, performance levels and optimizing becomes a backtest-overfitting exercise.

[FIGURE 18 HERE]

[FIGURE 19 HERE]

[FIGURE 20 HERE]

Figures 21 to 25 repeat the same process for $\mu = -10$ and $\tau$ that is progressively increased from 5 to 10, 25, 50 and 100. The same pattern arises, i.e. a rotated complementary to the case of positive long-run equilibrium.

# 5.- CONCLUSIONS

In this paper we have shown how to determine experimentally the optimal trading strategy associated with prices following a discrete Ornstein-Uhlenbeck process. Because the derivation of such trading strategy is not the result of a historical simulation, our procedure avoids the risks associated with backtest overfitting.

Depending on factors such as the frequency at which trading takes place, the holding period, etc., the time it takes to run our numerical procedure may be too lengthy. For that reason alone, it would be beneficial to count with a closed-form solution that computes the Sharpe ratio of every combination $\left(\underline{\pi_i}, \overline{\pi_i}\right)$, which we could then optimize analytically to determine the optimal $R_i^*$. In addition, although we can make sense of the experimental results presented in this paper, a closed-form representation of the solution would give us greater insight into what makes a particular solution $R_i^*$ optimal.

While in this paper we do not derive the closed-form solution to the optimal trading strategies problem, our experimental results seem to support the following OTR conjecture:

> *"Given a financial instrument's price characterized by a discrete O-U process, there is a unique optimal trading rule in terms of a combination of profit-taking and stop-loss that maximizes the rule's Sharpe ratio."*

We believe that solving this conjecture would have substantial economic value in a trading world where a few milliseconds separate winners from losers.

# APPENDICES

## A.1.- PYTHON IMPLEMENTATION OF OUR EXPERIMENTS

Snippet 1 provides an implementation in Python of the experiments conducted in this paper.

```python
#!/usr/bin/env python
# Profit-taking and stop-loss simulations
# On 20131003 by MLdP <lopezdeprado@lbl.gov>
import numpy as np
from random import gauss
from itertools import product
#----------------------------------------------------------------
def main():
    rPT=rSLm=np.linspace(0,10,21)
    count=0
    for prod_ in product([10,5,0,-5,-10],[5,10,25,50,100]):
        count+=1
        coeffs={'forecast':prod_[0],'hl':prod_[1],'sigma':1}
        output=batch(coeffs,nIter=1e5,maxHP=100,rPT=rPT,rSLm=rSLm)
    return output
#----------------------------------------------------------------
def batch(coeffs,nIter=1e5,maxHP=100,rPT=np.linspace(0,10,21), \
    rSLm=np.linspace(0,10,21),seed=0):
    phi,output1=2**(-1./coeffs['hl']),[]
    for comb_ in product(rPT,rSLm):
        output2=[]
        for iter_ in range(int(nIter)):
            p,hp,count=seed,0,0
            while True:
                p=(1-phi)*coeffs['forecast']+phi*p+coeffs['sigma']*gauss(0,1)
                cP=p-seed;hp+=1
                if cP>comb_[0] or cP<-comb_[1] or hp>maxHP:
                    output2.append(cP)
                    break
        mean,std=np.mean(output2),np.std(output2)
        print comb_[0],comb_[1],mean,std,mean/std
        output1.append((comb_[0],comb_[1],mean,std,mean/std))
    return output1
```

*Snippet 1 – Python code for the determination of OTRs*

The subroutine *batch(\*arg)* estimates the Sharpe ratios for each node of the mesh $\left(\underline{\pi_i}, \overline{\pi_i}\right)$ given some input parameters $\{\hat{\sigma}, \hat{\varphi}\}$ and initial conditions $\{P_{i,0}, E_0[P_{i,T_i}]\}$. *batch(\*arg)* is called by *main()*, which passes alternative values of $\hat{\varphi}$ and $E_0[P_{i,T_i}]$. A path is discontinued after 100 steps, thus values of $\hat{\varphi}$ are explored such that $\tau \in \{5,10,25,50,100\}$. As forecasted prices, we have tried $E_0[P_{i,T_i}] \in \{-10,-5,0,5,10\}$. We have fixed $P_{i,0} = 0$, since it is the distance $\left(P_{i,t-1} - E_0[P_{i,T_i}]\right)$ that drives the convergence, not particular absolute price levels. Also without loss of generality, in all simulations we have used $\hat{\sigma} = 1$.

**TABLES**

| ID | TableName | Forecast | HL | Sigma | maxHP |
|----|-----------|----------|-----|-------|-------|
| 1  | Table_1   | 0        | 5   | 1     | 100   |
| 2  | Table_2   | 0        | 10  | 1     | 100   |
| 3  | Table_3   | 0        | 25  | 1     | 100   |
| 4  | Table_4   | 0        | 50  | 1     | 100   |
| 5  | Table_5   | 0        | 100 | 1     | 100   |
| 6  | Table_6   | 5        | 5   | 1     | 100   |
| 7  | Table_7   | 5        | 10  | 1     | 100   |
| 8  | Table_8   | 5        | 25  | 1     | 100   |
| 9  | Table_9   | 5        | 50  | 1     | 100   |
| 10 | Table_10  | 5        | 100 | 1     | 100   |
| 11 | Table_11  | 10       | 5   | 1     | 100   |
| 12 | Table_12  | 10       | 10  | 1     | 100   |
| 13 | Table_13  | 10       | 25  | 1     | 100   |
| 14 | Table_14  | 10       | 50  | 1     | 100   |
| 15 | Table_15  | 10       | 100 | 1     | 100   |
| 16 | Table_16  | -5       | 5   | 1     | 100   |
| 17 | Table_17  | -5       | 10  | 1     | 100   |
| 18 | Table_18  | -5       | 25  | 1     | 100   |
| 19 | Table_19  | -5       | 50  | 1     | 100   |
| 20 | Table_20  | -5       | 100 | 1     | 100   |
| 21 | Table_21  | -10      | 5   | 1     | 100   |
| 22 | Table_22  | -10      | 10  | 1     | 100   |
| 23 | Table_23  | -10      | 25  | 1     | 100   |
| 24 | Table_24  | -10      | 50  | 1     | 100   |
| 25 | Table_25  | -10      | 100 | 1     | 100   |

*Table 1 – Input parameter combinations used in the simulations*

# FIGURES

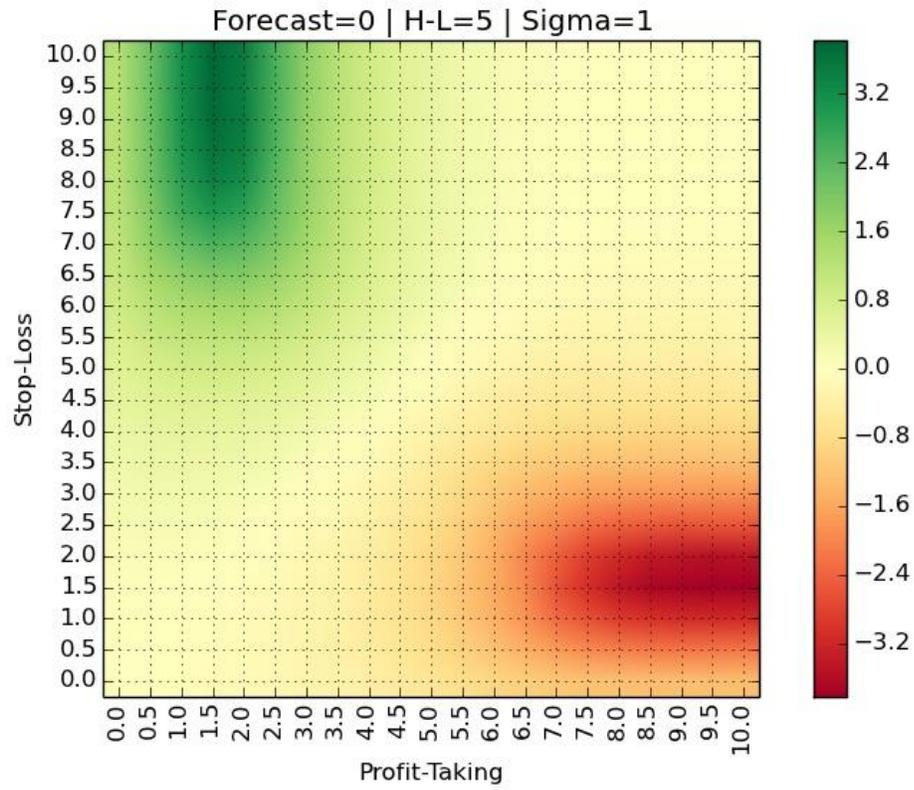

*Figure 1 – Heat-map for $\{\mu, \tau, \sigma\} = \{0,5,1\}$*

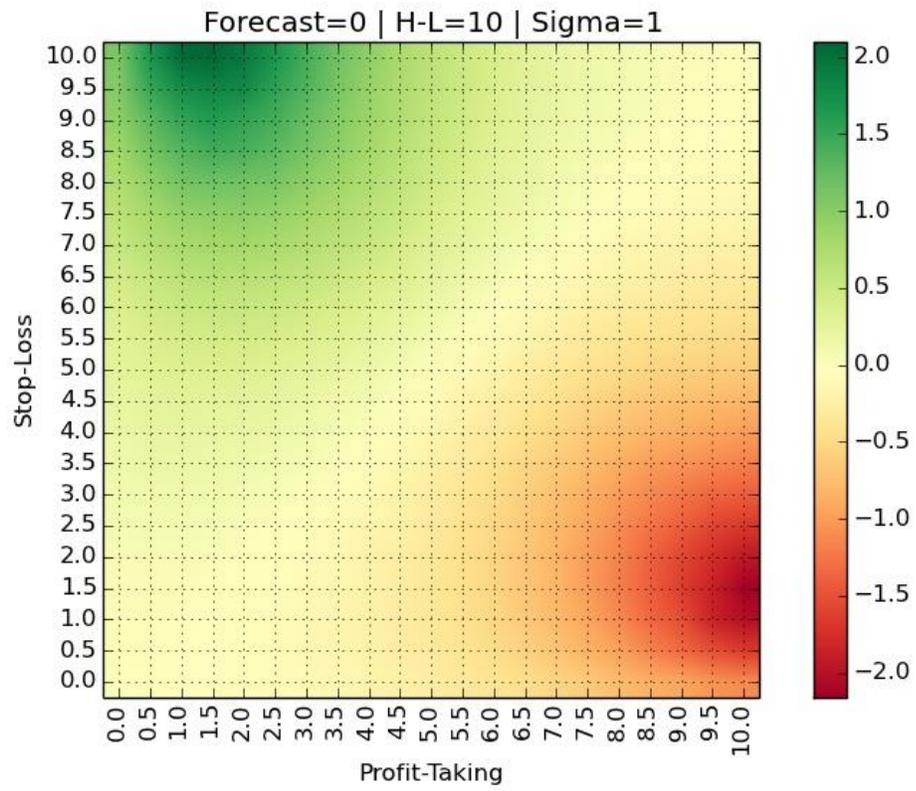

Figure 2 – *Heat-map for* $\{\mu, \tau, \sigma\} = \{0,10,1\}$

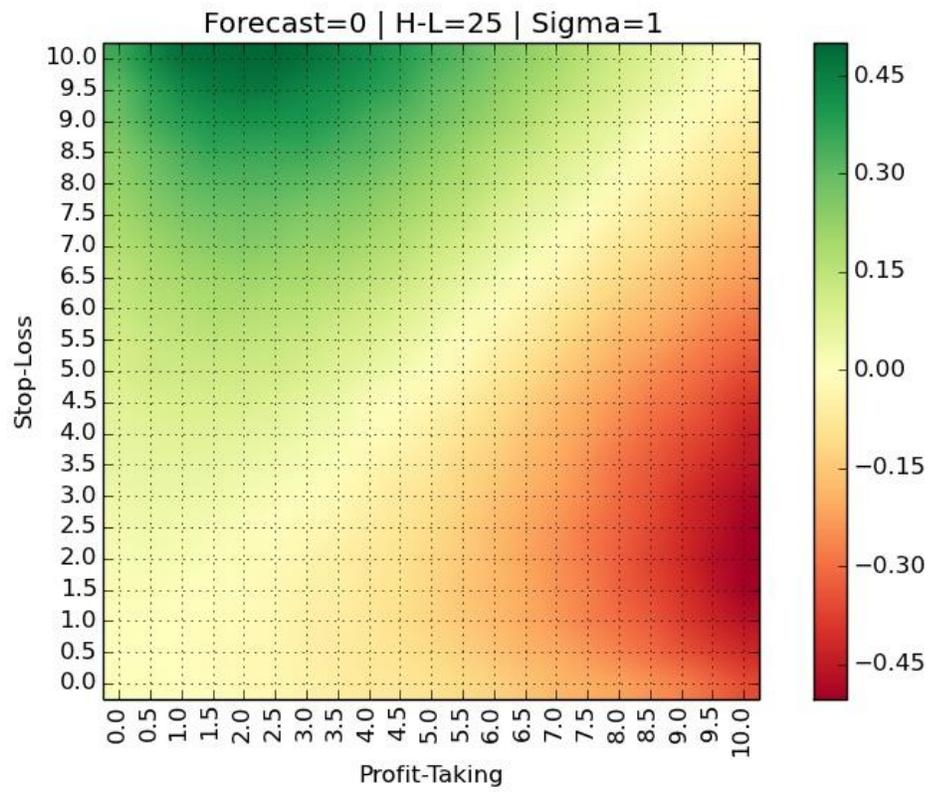

Figure 3 – *Heat-map for* $\{\mu, \tau, \sigma\} = \{0,25,1\}$

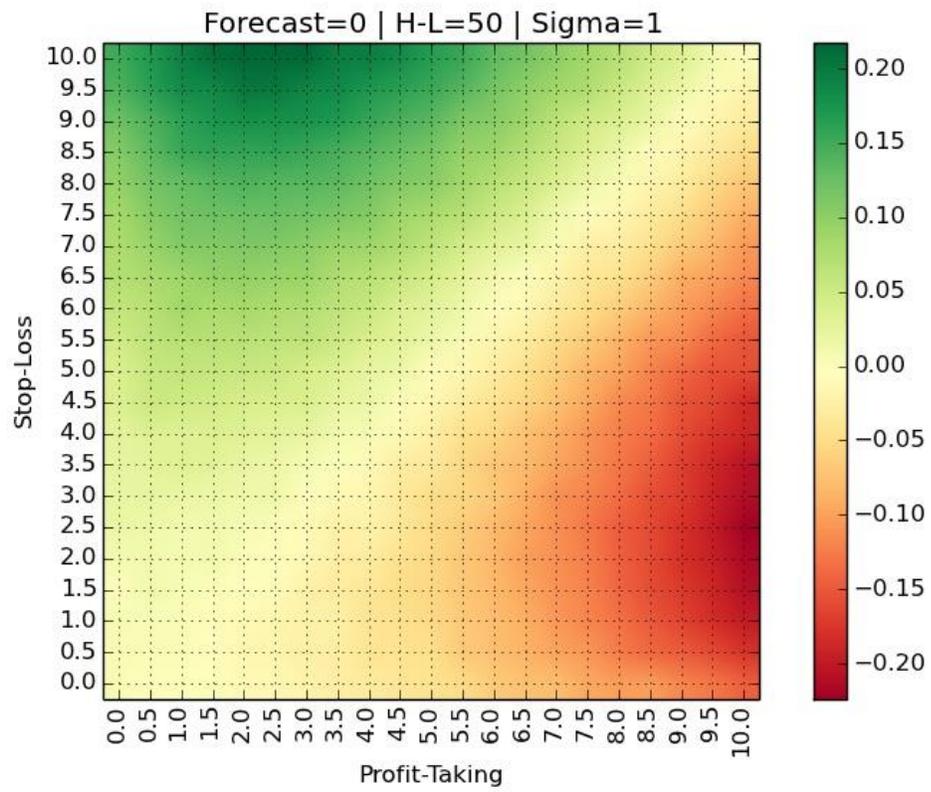

Figure 4 – *Heat-map for* $\{\mu, \tau, \sigma\} = \{0, 50, 1\}$

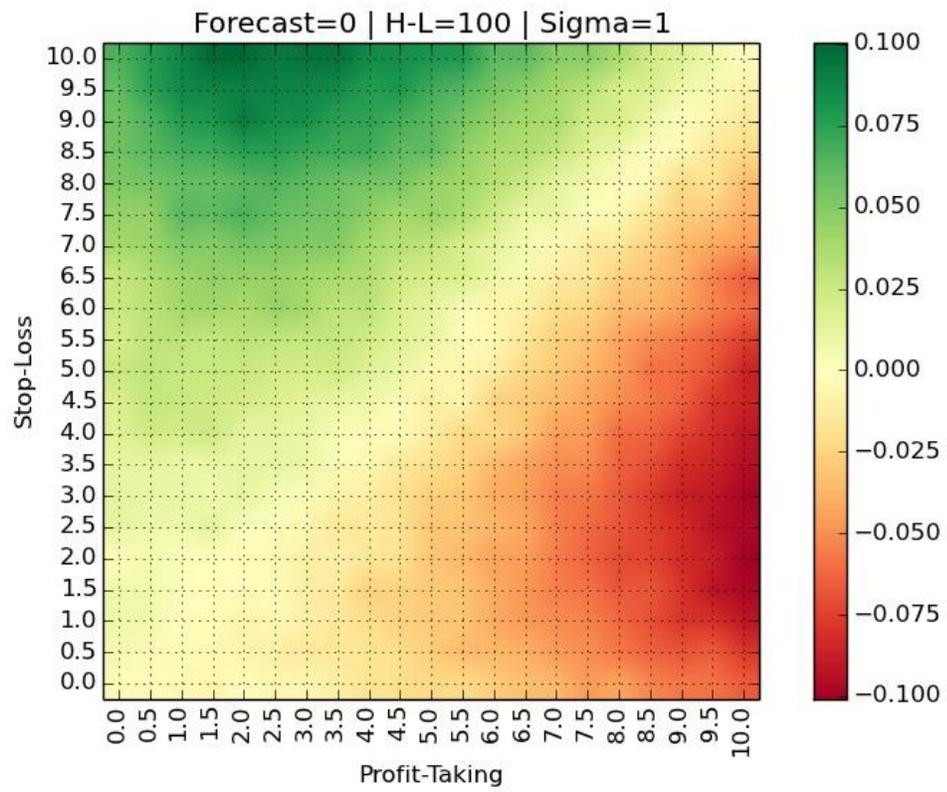

Figure 5 – *Heat-map for* $\{\mu, \tau, \sigma\} = \{0,100,1\}$

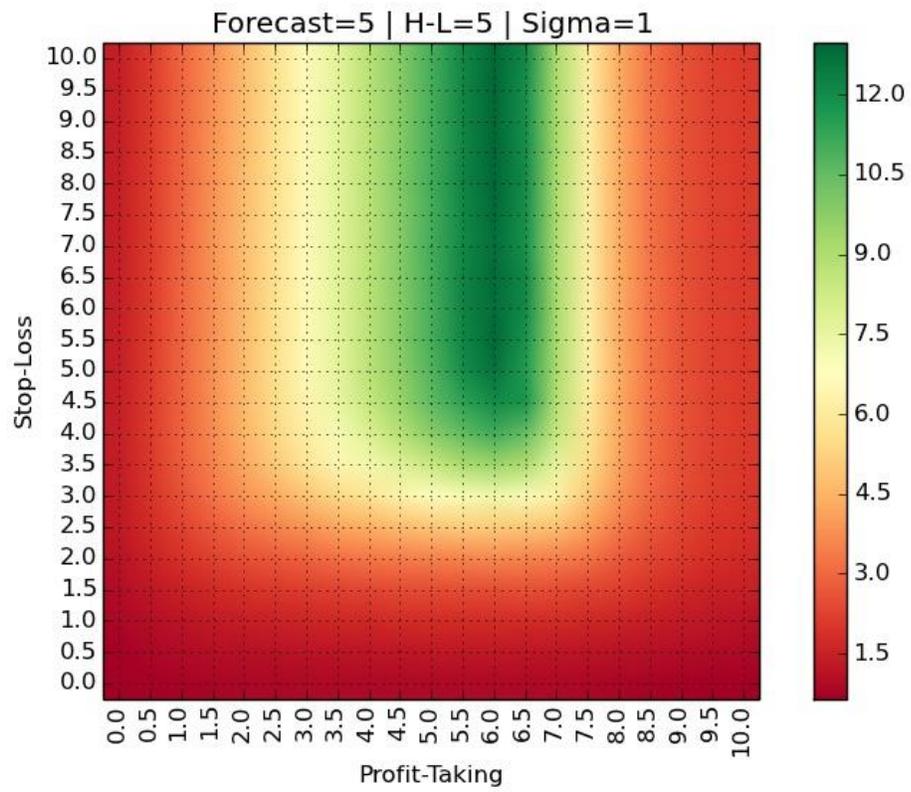

Figure 6 – *Heat-map for* $\{\mu, \tau, \sigma\} = \{5, 5, 1\}$

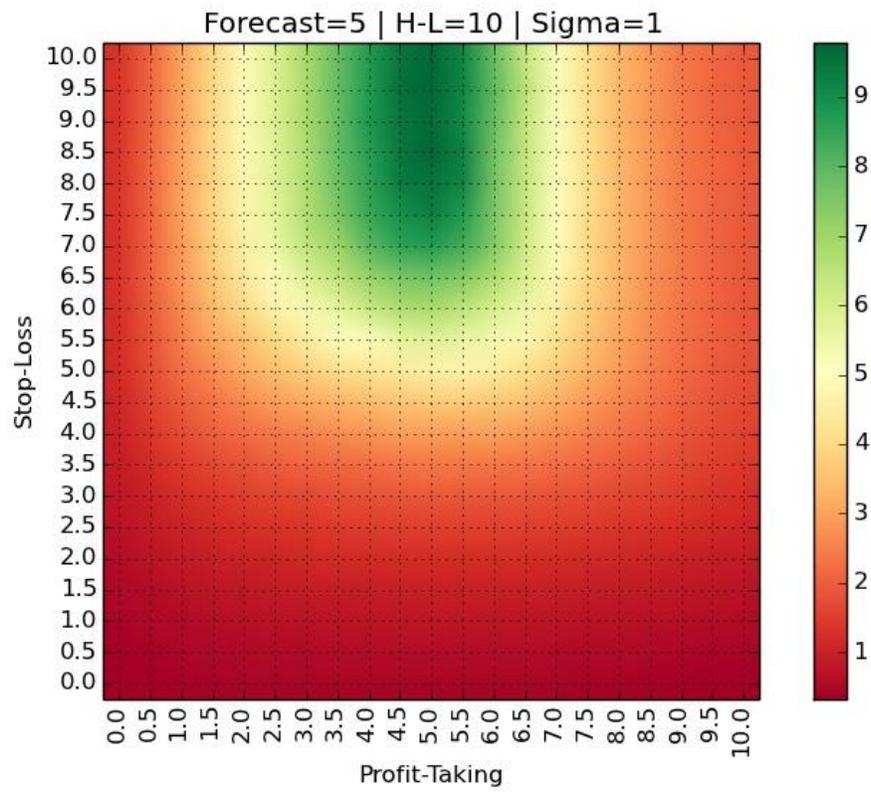

Figure 7 – *Heat-map for* $\{\mu, \tau, \sigma\} = \{5, 10, 1\}$

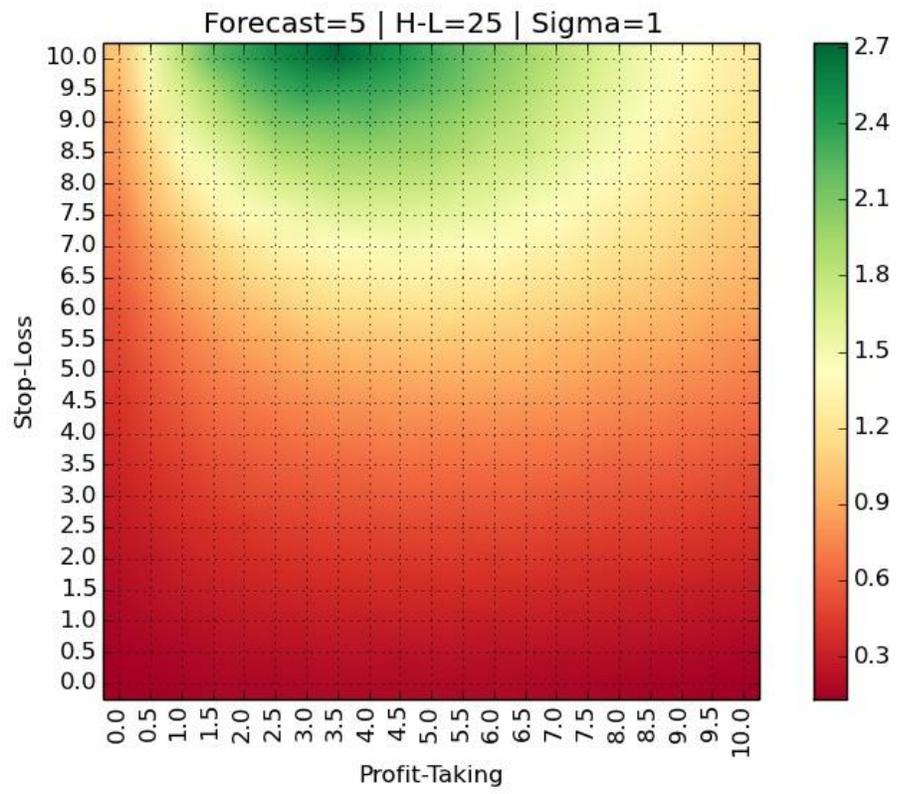

Figure 8 – *Heat-map for* $\{\mu, \tau, \sigma\} = \{5, 25, 1\}$

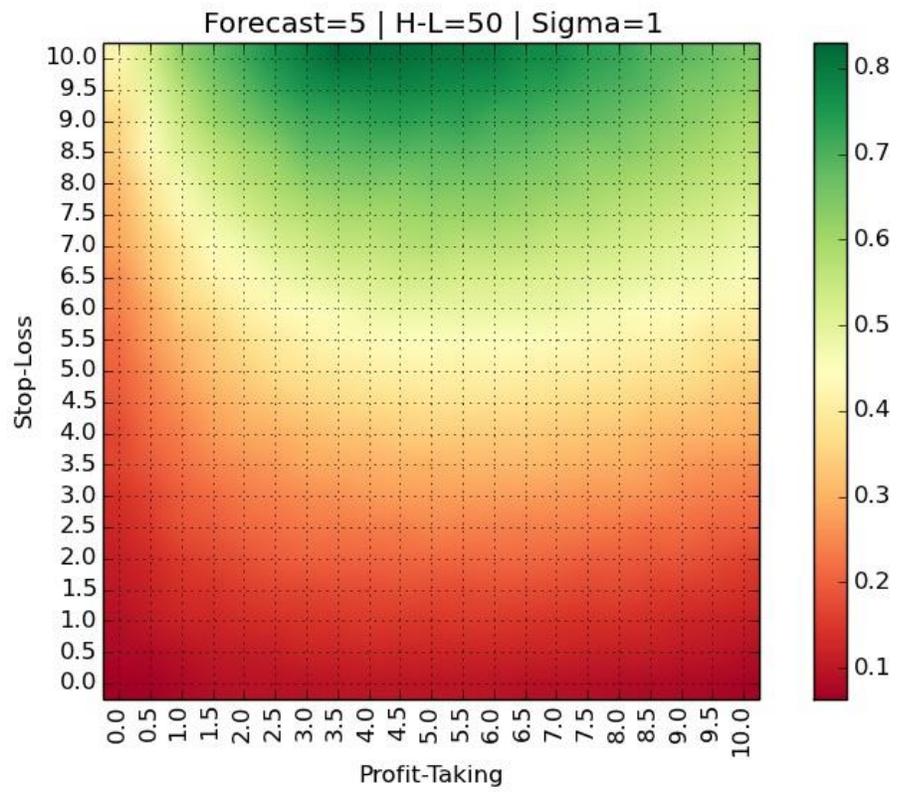

Figure 9 – *Heat-map for* $\{\mu, \tau, \sigma\} = \{5, 50, 1\}$

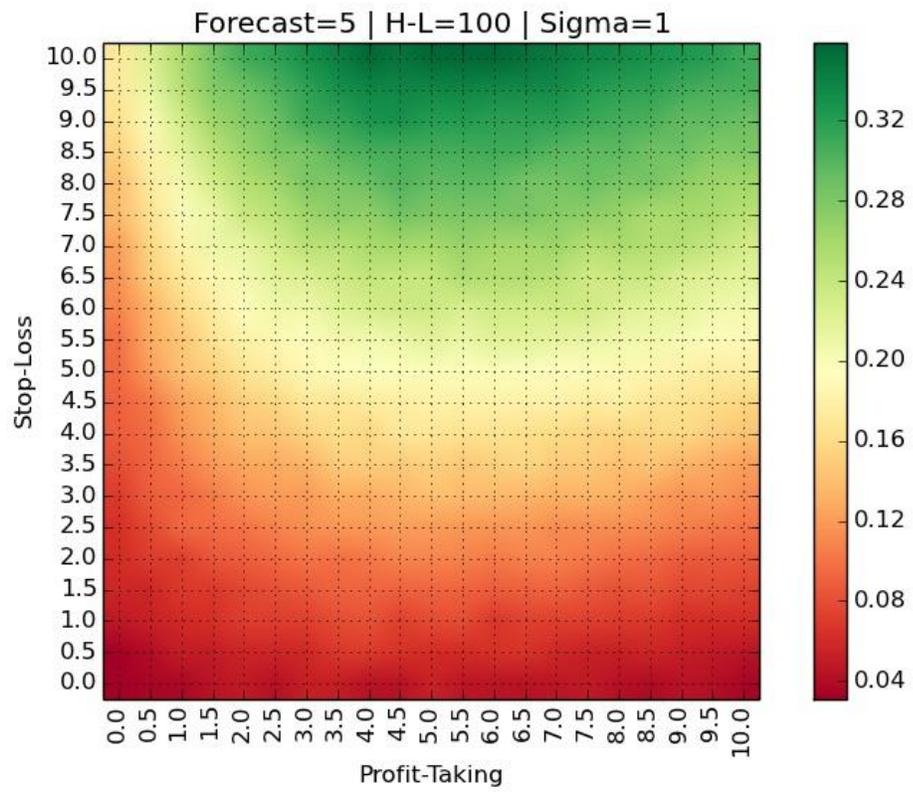

Figure 10 – *Heat-map for* $\{\mu, \tau, \sigma\} = \{5, 100, 1\}$

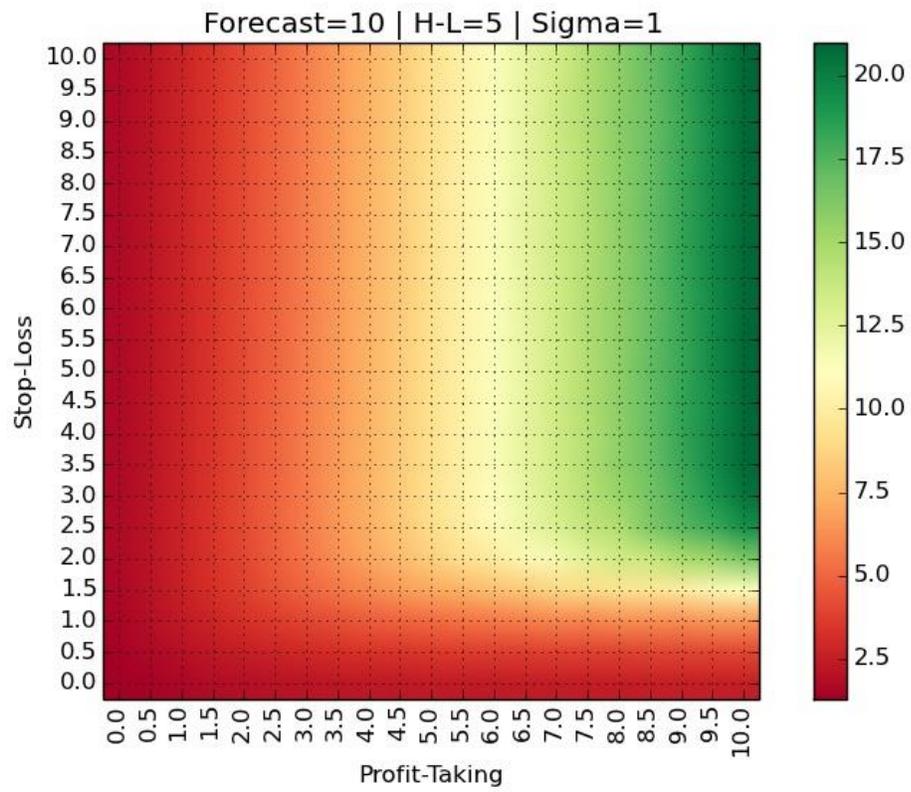

Figure 11 – *Heat-map for* $\{\mu, \tau, \sigma\} = \{10, 5, 1\}$

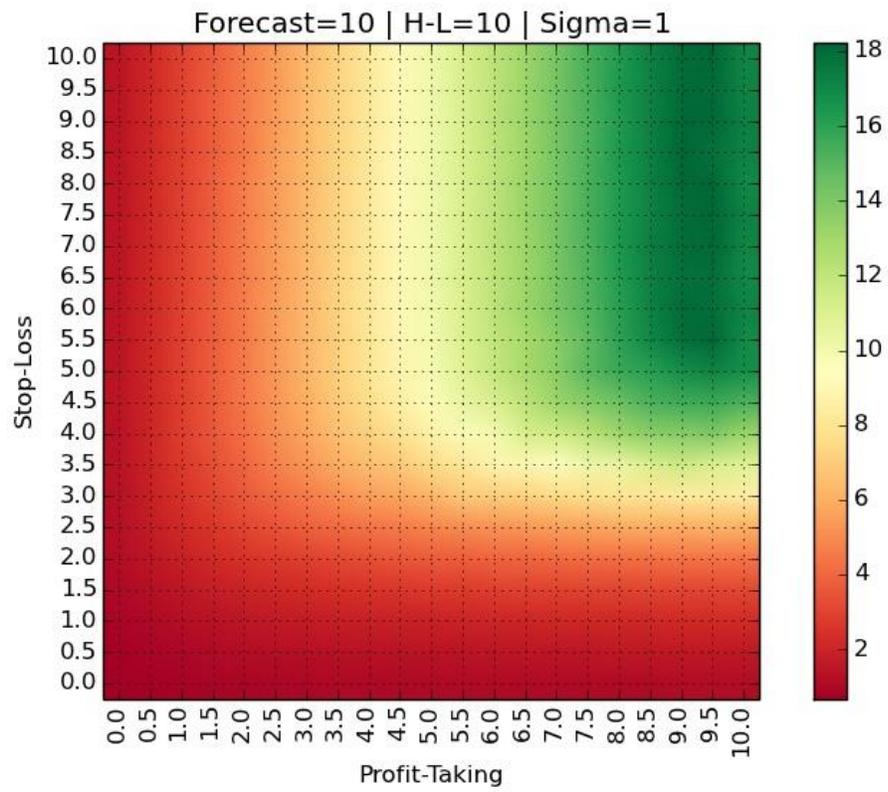

Figure 12 – *Heat-map for* $\{\mu, \tau, \sigma\} = \{10,10,1\}$

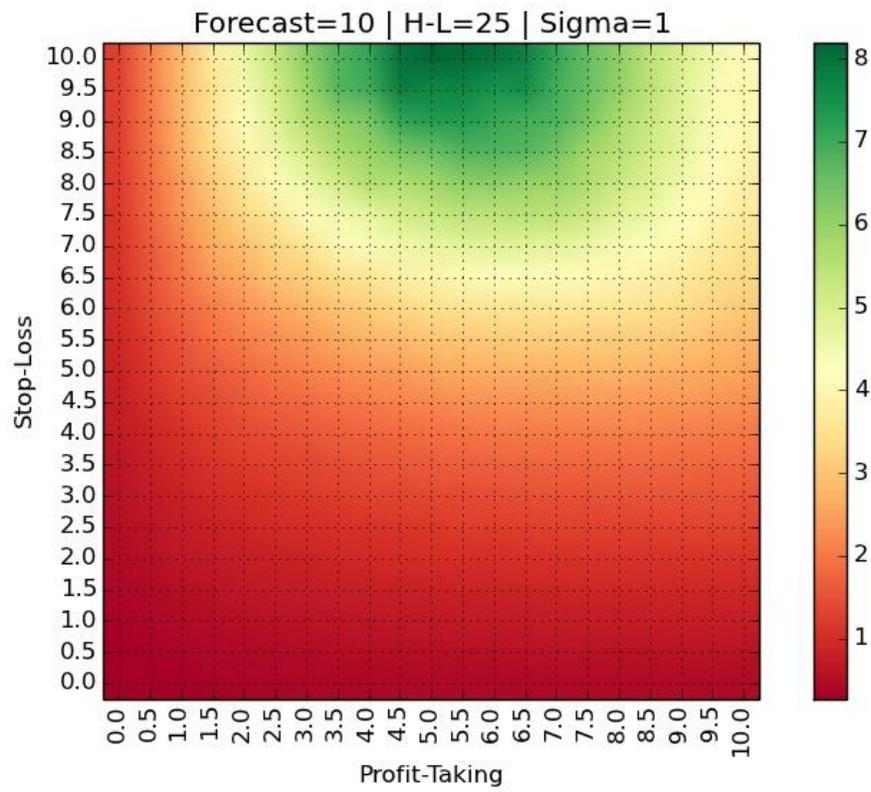

Figure 13 – *Heat-map for* $\{\mu, \tau, \sigma\} = \{10, 25, 1\}$

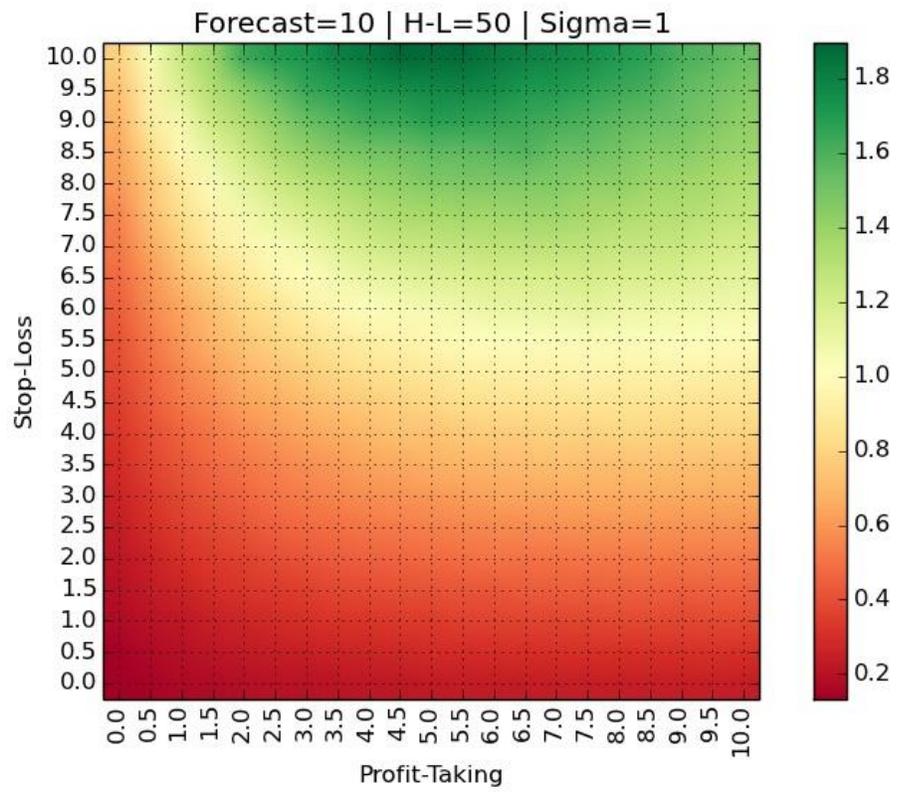

Figure 14 – *Heat-map for* $\{\mu, \tau, \sigma\} = \{10, 50, 1\}$

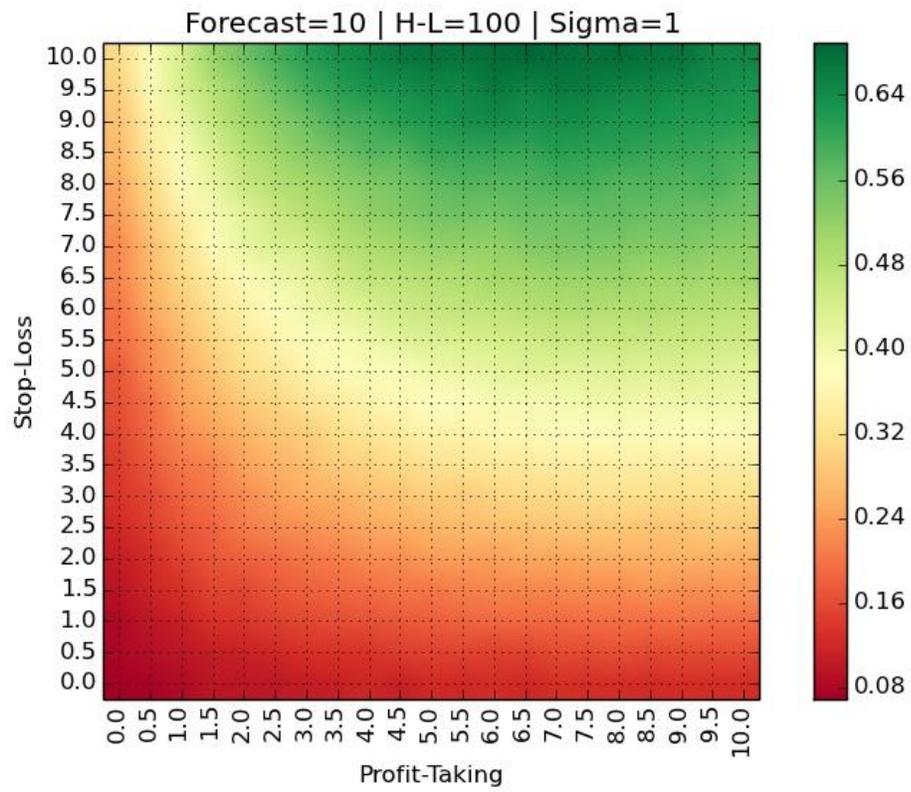

Figure 15 – *Heat-map for* $\{\mu, \tau, \sigma\} = \{10, 100, 1\}$

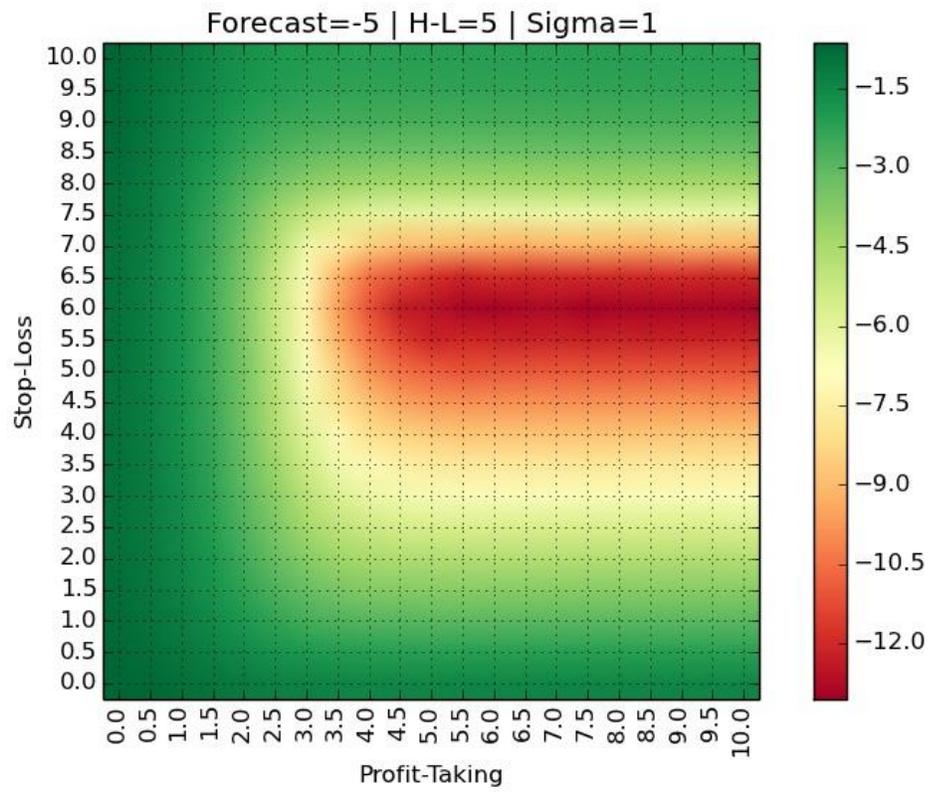

Figure 16 – *Heat-map for* $\{\mu, \tau, \sigma\} = \{-5,5,1\}$

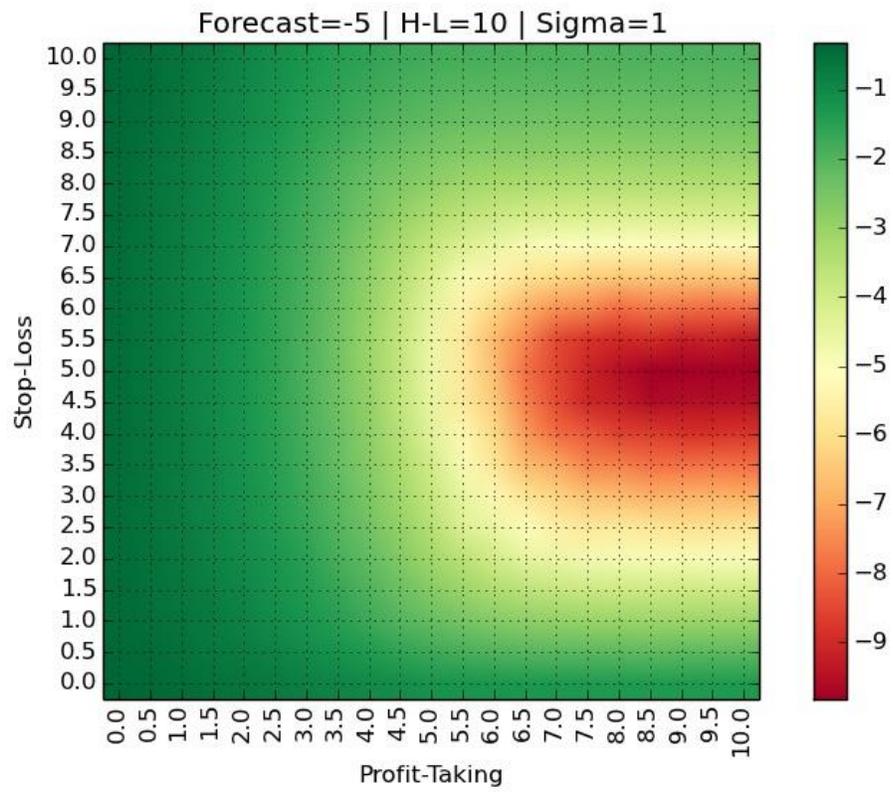

Figure 17 – *Heat-map for* $\{\mu, \tau, \sigma\} = \{-5, 10, 1\}$

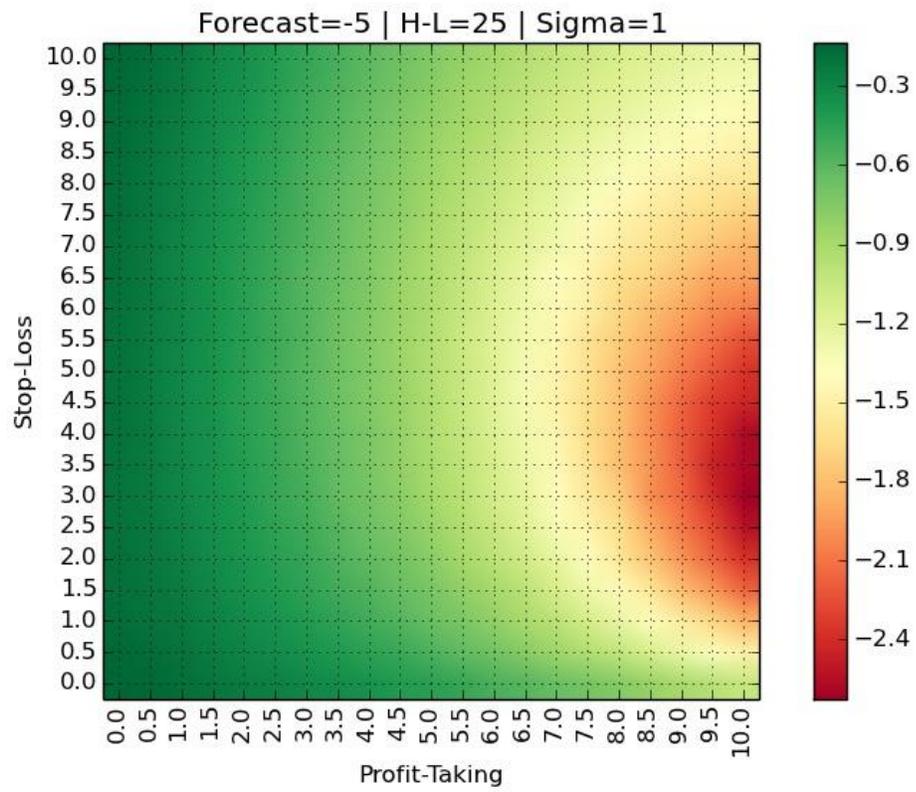

Figure 18 – *Heat-map for* $\{\mu, \tau, \sigma\} = \{-5, 25, 1\}$

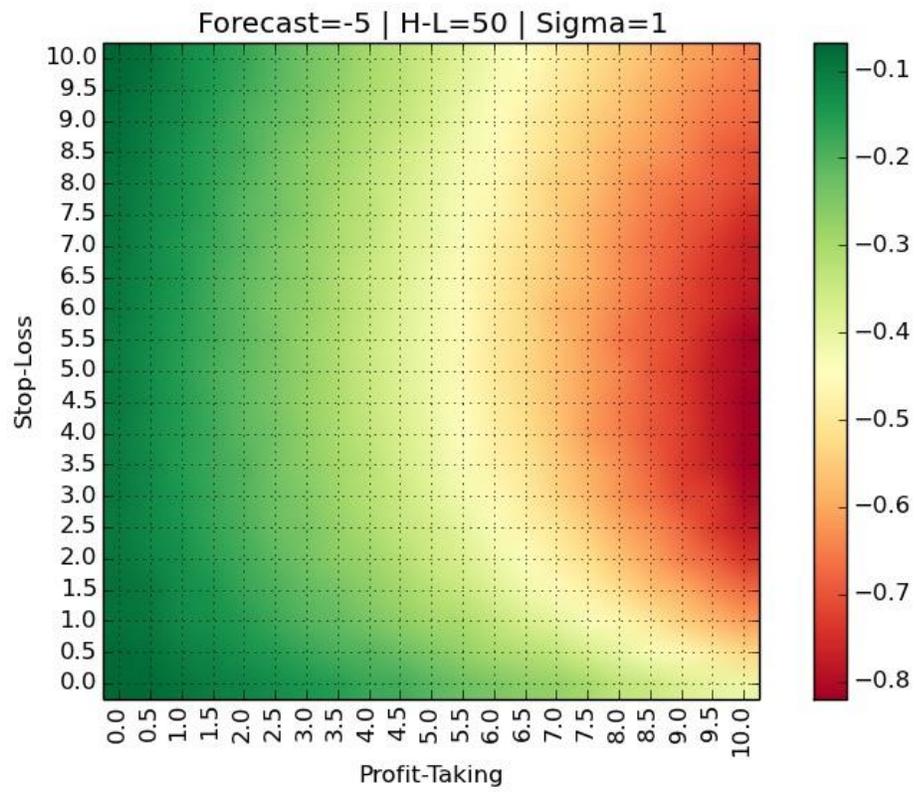

Figure 19 – *Heat-map for* $\{\mu, \tau, \sigma\} = \{-5, 50, 1\}$

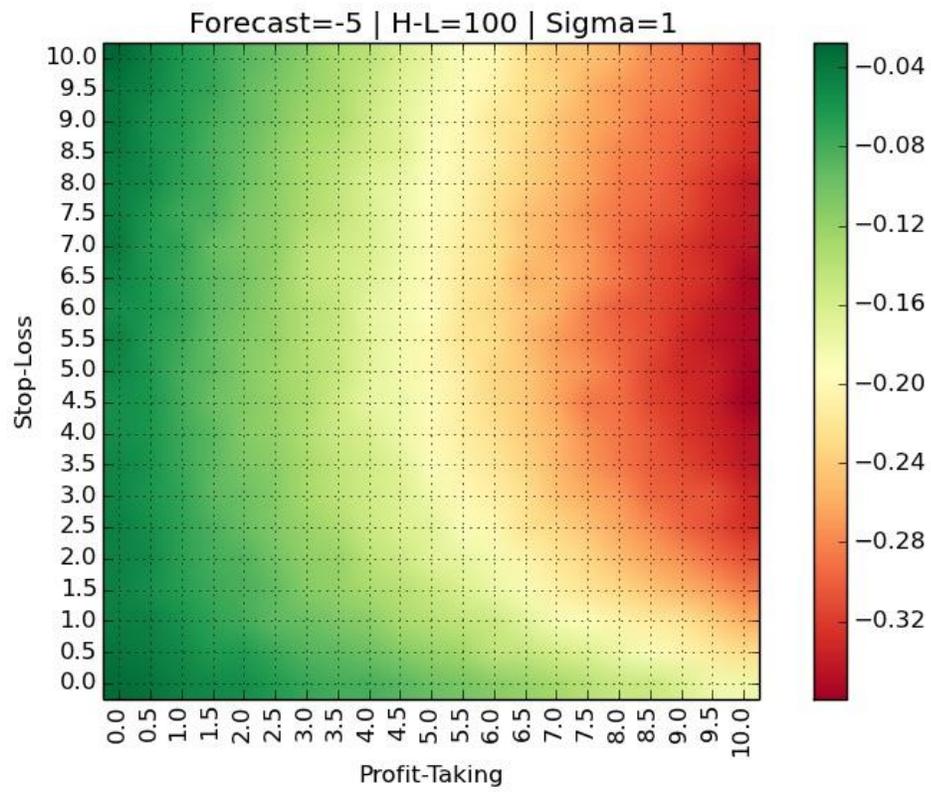

Figure 20 – *Heat-map for* $\{\mu, \tau, \sigma\} = \{-5, 100, 1\}$

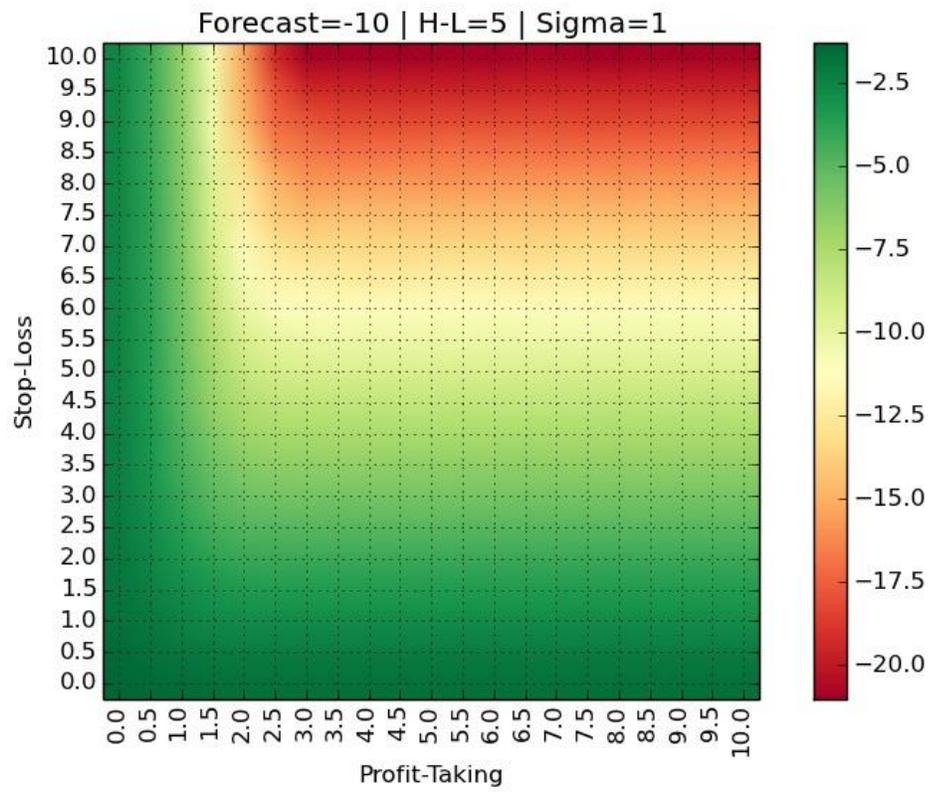

Figure 21 – *Heat-map for* $\{\mu, \tau, \sigma\} = \{-10, 5, 1\}$

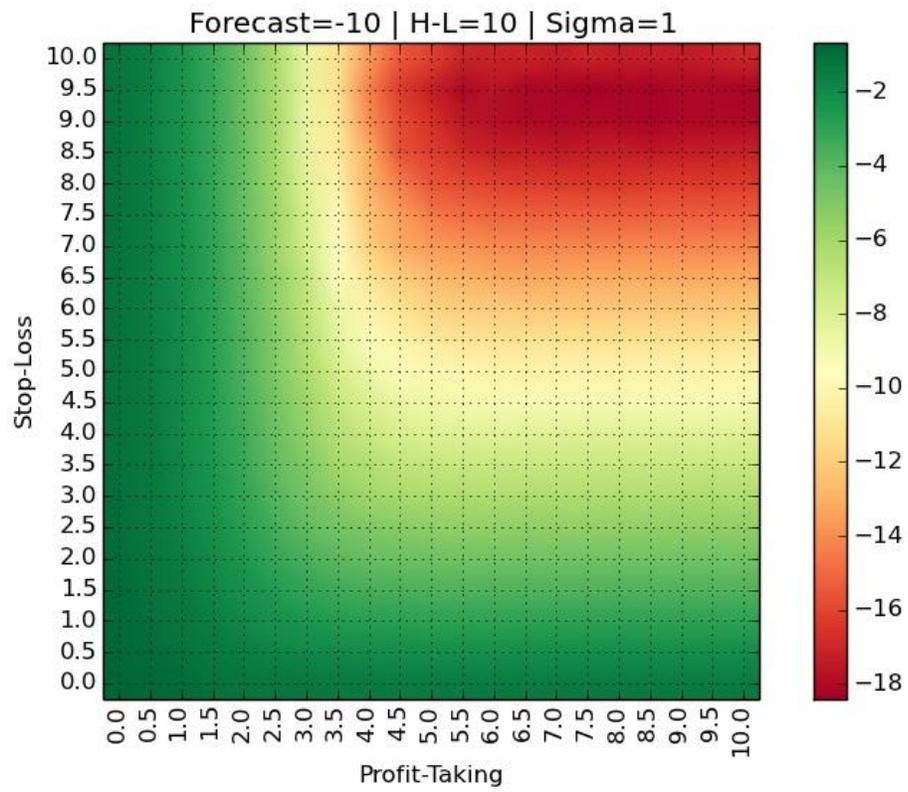

Figure 22 – *Heat-map for* $\{\mu, \tau, \sigma\} = \{-10,10,1\}$

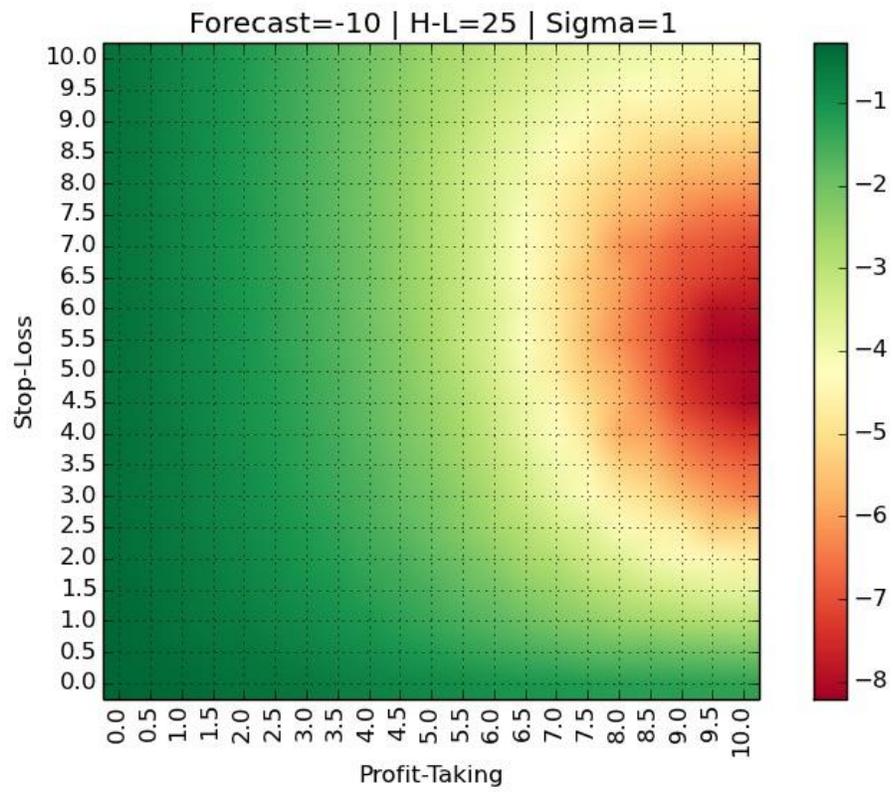

Figure 23 – *Heat-map for* $\{\mu, \tau, \sigma\} = \{-10, 25, 1\}$

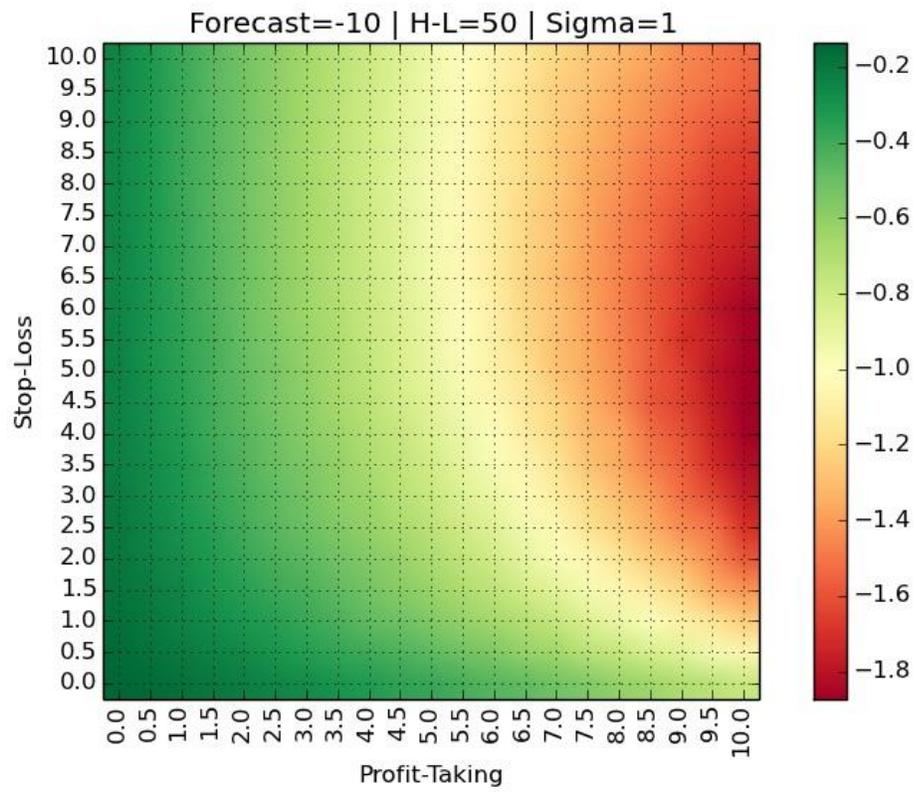

Figure 24 – *Heat-map for* $\{\mu, \tau, \sigma\} = \{-10, 50, 1\}$

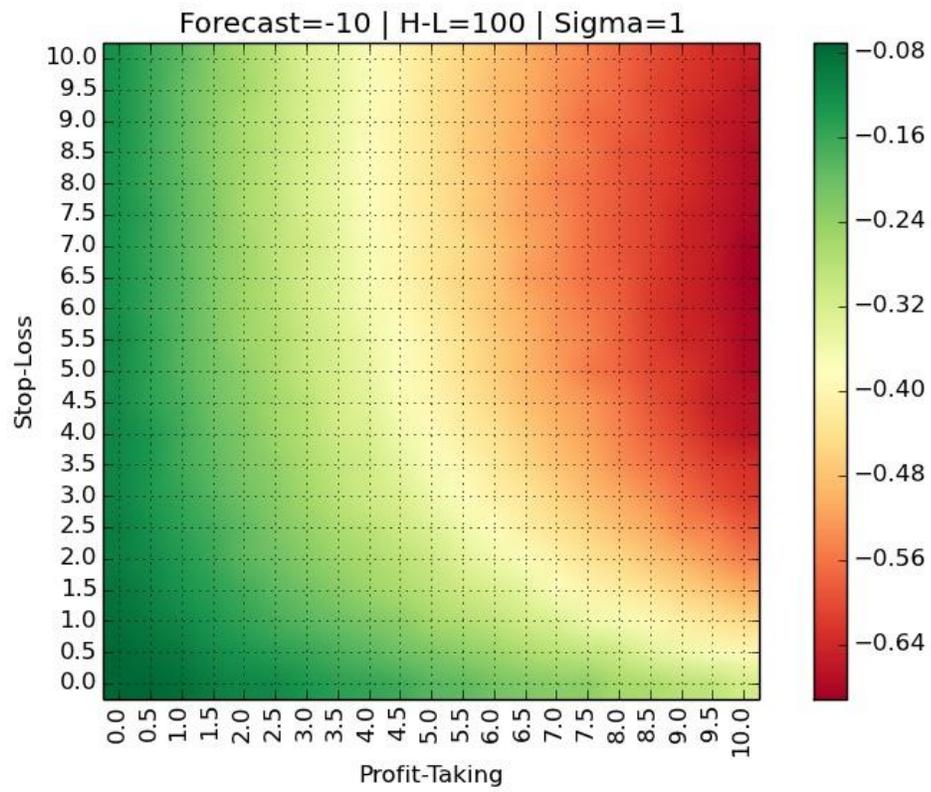

Figure 25 – *Heat-map for* $\{\mu, \tau, \sigma\} = \{-10,100,1\}$